# Surface Plasmon Polariton Self-Focusing by Ponderomotive Forces

**Pavel Ginzburg, Alex Hayat, Eyal Feigenbaum, Nikolai Berkovitch and Meir Orenstein**

*Department of Electrical Engineering, Technion, Haifa 32000, Israel*

meiro@ee.technion.ac.il

Nonlinear properties of Surface Plasmon Polaritons stemming from the inherent electron plasma nonlinearity of the metal layers are investigated. A fluid-mechanics plasma model is used to describe the electron motion in metals. The nonlinear ponderomotive force repels the electrons from the high field intensity region, effectively reducing the local plasma frequency and the corresponding real part of the refractive index results in Kerr like self-focusing. The field confinement to the low carrier density region also assists also in reducing the propagation losses, which usually inhibit practical nanoplasmonic circuits. Surface plasmon self focusing and nonlinear spectrum modifications, predicted by our model, are demonstrated by preliminary experiments.

Sub-wavelength Surface Plasmon Polariton (SPP) optics show promise for the realization of nanometer-size photonic integrated circuits for applications such as optical interconnects, signal processing and nanosensing. The basic SPP phenomenon is guiding of light by a single dielectric/metal interface [1]. Two-dimensional beam self focusing based on Kerr nonlinearity, which occurs in specific media and for high light intensities, is well known. Light guiding on a single interface due to the nonlinear response have been studied intensively [2]. Most of such reports consider the nonlinearity of at least one of the dielectric layers and derive appropriate soliton conditions. On the other hand, nonlinear plasma response to intense laser beams, such as self-focusing and guiding, was studied in beam-induced fusion and ionospheric research. The noble metals e.g. gold and silver may be treated as cold-electron plasma immersed in a lattice of positively charged ions in the visible and the infrared frequencies range.

Here we study the nonlinear properties of SPP, stemming from the inherent electron plasma nonlinearity of the metal layers. Other possible nonlinear mechanisms heating and collisions were ruled out here due to very short pulse excitation and to the exceptionally high validity of the collision less Drude model with the linear results of SPP propagation. The fluid model is commonly used in plasma physics for electron motion in metals [3]. The nonlinear force due to electromagnetic wave propagation within the plasma – the ponderomotive (or Miller) force, repels the electrons from the region of high field intensity, effectively reducing the local plasma frequency and, hence, reducing the (absolute value) of the electric permittivity constant of the metal. This reduction enhances the effective index of the high intensity peak of single-interface SPP and, similarly to the Kerr effect, results in beam self-confinement. A qualitative explanation may be given by the effective index model - electron escape from the region of high intensity leads to enhancement of the real part of the local refractive index of SPP $n=\sqrt{\varepsilon_m \varepsilon_d /(\varepsilon_m + \varepsilon_d)}$ , and hence to self focusing. Another interesting feature is the confinement of the field to the region of low carrier density that leads to reduction of the propagation losses.

The nonlinear plasma response in expressed by [4, 5]:

$$\varepsilon_m(E,\omega) = 1 - \left(\frac{\omega_p}{\omega}\right)^2 exp\left(-\alpha \langle E^2 \rangle\right)$$

$$\alpha = \frac{e^2}{2m_e \omega^2 (T_e + T_i)}$$

(1)

where $\omega_p$ is the plasma frequency (without an external field), $\omega$ is the light frequency, e and $m_e$ are the electron charge and mass respectively, $T_e$ and $T_i$ are the temperatures of electrons and ions in the lattice and E is the light field. $<E^2>$ is the time independent average intensity.

The resulting wave equation is:

$$\begin{cases} \nabla \times \nabla \times \bar{E} - k_0^2 \varepsilon_M \bar{E} = 0, metal \\ \nabla^2 E + \mu \varepsilon_d \omega^2 E = 0, \quad dielectric \end{cases}$$

(2)

for the nonlinear region (metal substrate) and linear dielectric cladding. $\varepsilon_d$ is the linear dielectric constant of the cladding and $\varepsilon_m$, defined by Eq. 1, is the nonlinear dielectric constant of metal substrate. The real part of $\varepsilon_m$ is negative for visible and infrared light, hence, even at the linear regime guided transverse magnetic (TM) waves are supported by the single interface (Fig. 1(a)).

For TM modes $\partial_z E_z = \partial_x E_x$, thus $\nabla \times \nabla \times \bar{E} = -\nabla^2 E_x \hat{x} - \nabla^2 E_z \hat{z}$, and the wave equation become:

$$(\nabla^2 E_x \hat{x} + \nabla^2 E_z \hat{z}) + k_0^2 \varepsilon_M \exp\left\{-\alpha \left(|E_x|^2 + |E_z|^2\right)\right\} (E_x \hat{x} + E_z \hat{z}) = 0$$

(3)

Assuming small non-linearity:

$$\exp\left\{-\alpha\left(|E_x|^2 + |E_z|^2\right)\right\} \sim 1 - \alpha\left(|E_x|^2 + |E_z|^2\right) \quad (4)$$

The wave equation is:

$$(\nabla^2 E_x \hat{x} + \nabla^2 E_z \hat{z}) + k^2 \vec{E} - k_0^2 \varepsilon_M \alpha\left(|\vec{E}|^2\right)\vec{E} = 0$$
$$\vec{E} \triangleq (E_x \hat{x} + E_z \hat{z}) \quad (5)$$

The field is written as [6] $\vec{E} = \vec{E}_0(x) A(z, y) \exp\{j\beta z\}$, where $E_0$ is the TM field distribution for vanishing nonlinearity ($\vec{E}_0 = E_x \hat{x} + E_z \hat{z}$), $A$ is a slowly varying envelope, and $\beta$ is the $z$ propagation constant. By employing a paraxial approximation for $z$ propagation (namely omitting $A_{zz}$), multiplying it by $\vec{E}_0^*$ and averaging over x yields a scalar wave equation for the amplitude $A$:

$$j2\beta A_z + \left[(k^2 - \beta^2) + I_2/I\right] A + A_{yy} - \alpha k_0^2 \varepsilon_M (I_3/I) |A|^2 A = 0 \quad (6)$$

Replacing the slowly varying amplitude { a=A·exp{-j(k²-β²+I₂/I)/(2β)z)}} yields:

$$j a_z + (1/2\beta) a_{yy} - (\alpha k_0^2 \varepsilon_M /2\beta)(I_3/I)|A|^2 A = 0 \quad (7)$$

α is a positive constant, which together with the negative metal dielectric constant yields bright soliton (rather than dark soliton), with peak amplitude η and width Δy:

$$\Delta y \cdot \eta = \sqrt{I/(\alpha|\varepsilon_M|I_3)} \, (\lambda_0/2\pi) \quad (8)$$

Where the nonlinear coefficients are modified by terms of the vertical mode: $I_3$ is the vertically integrated squared intensity in the metal and I is the total integrated intensity:

$$I = \int_{-\infty}^{+\infty} \left|\vec{E}_0\right|^2 dx; \quad I_3 = \int_{metal} \left|\vec{E}_0\right|^4 dx \qquad (9)$$

Since the single surface plasmon polariton mode is known in close form, so are the constants of the nonlinear Schrödinger equation.

The nonlinear coefficient α (Eq. 1) for gold substrate at room temperature is about $3 \cdot 10^{-18}$ [$m^2/V^2$], while the nonlinear refractive index, derived from Eq. 6 is $n_2 \sim 10^{-18}$ [$cm^2/V^2$], which is comparable to highly nonlinear glasses [7] promising significant nonlinear effects, due to the extreme field enhancement in plasmonic devices. The schematic plot of a calculated soliton shape is presented on Fig. 1(b):

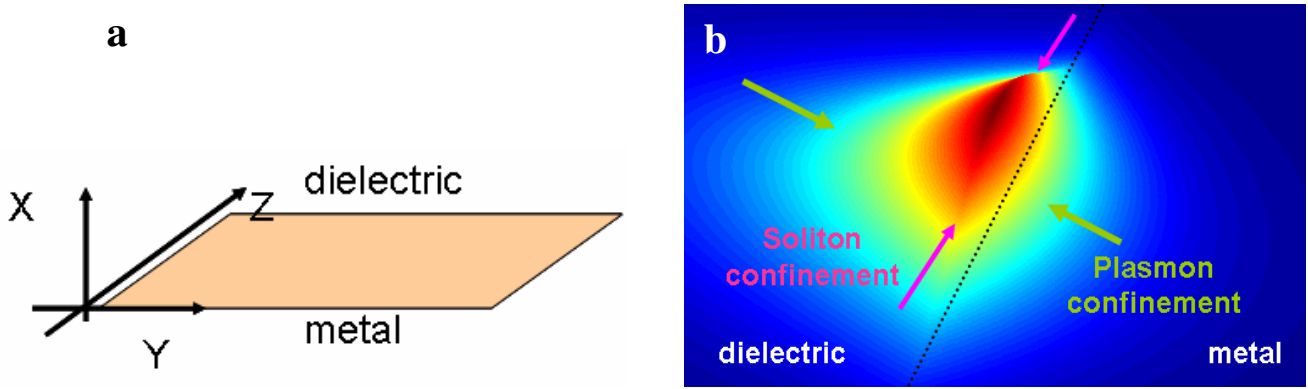

**Fig.1.** (a) Basic scheme of the structure; (b) the field profile – green arrows represent the confinement by metal/dielectric interface; magenta arrows indicate the soliton confinement.

In order to demonstrate experimentally this self-focusing effects we employed a measurement set up based on Near-Field Scanning Optical Microscope (NSOM). A flat and wide 70nm thick gold film was end fire excited by a lensed single mode fiber. The exciting optical pulses of ~1 psec duration at λ~1.5μm were generated using an optical parametric oscillator pumped by a mode-locked Ti:Sapphire femto-second laser at 810nm. The average pump power used was ~30mW and the pump polarization was TM. A digital attenuator, incorporated in the setup was employed to change the input intensity.

We measured the mode profile as the function of input attenuation (Fig. 2). The power-dependent mode width, decreasing with enhanced input power, evidently shows indications for nonlinear self-focusing. It should be emphasized that measurments using NSOM are limited in their dynamic range – thus we could not employ a larger span of input intensities.

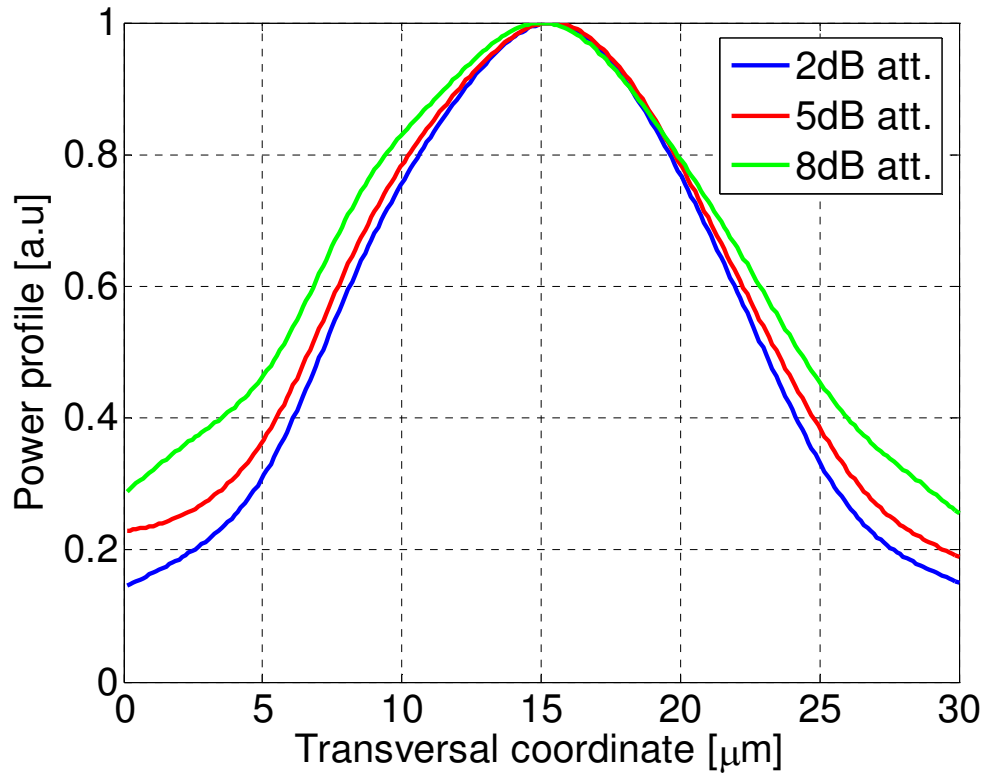

**Fig.2. Normalized intensity-dependent profile of the surface mode**

Additionally, we observed a spectral signature of a nonlinear effect - the intensity spectrum dependence of plasmonic pulses, now in a v-groove waveguide (Fig. 3 inset). The waveguide was optically pumped in a configuration described above, while the output power was fiber-coupled into an ANDO spectrum analyzer. Our measurements (Fig. 3) show clear indications for a non linear spectral shrinkage as a function of the input power.

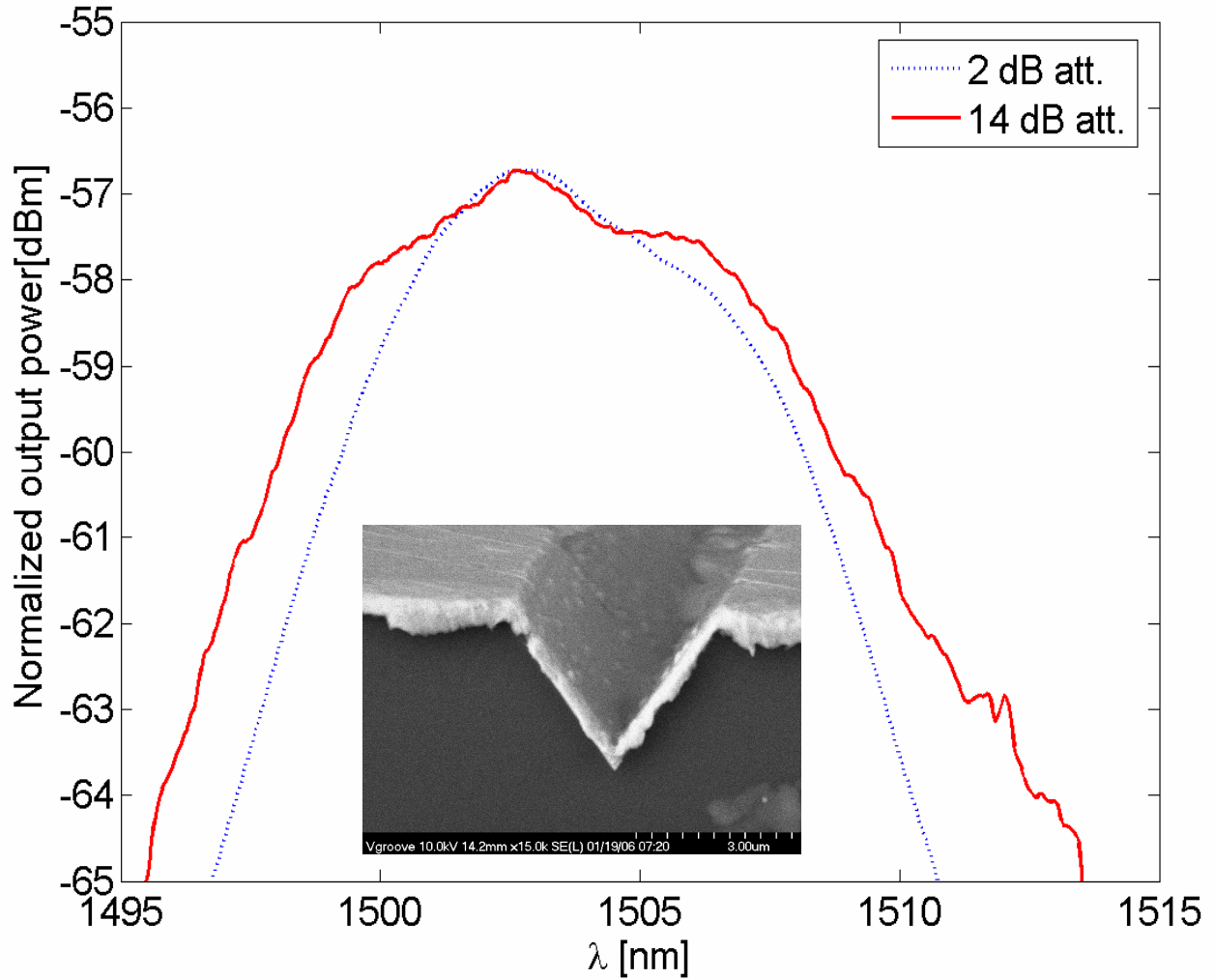

**Fig.3. Intensity dependent spectrum of v-groove plasmonic waveguide.**

**The inset is a Scanning Electron Microscope image of v-groove waveguide.**

In conclusion, we theoretically predict the existence of nonlinear surface plasmon waves due to ponderomotive forces on a metal substrate. Conditions for bright spatial soliton propagation were derived. Experimental results show indication for power dependence of the spatial surface plasmon polariton modal width and spectrum modifications.